Mantle-derived magmatic anhydrite in ultramafic rocks reveals deep mantle oxidation and mega-mineralization after Palaeozoic Oxygenation Event

First author: Yuegao Liu

# Abstract

Magmatic anhydrite has long been regarded as diagnostic of oxidized intermediate–felsic magmas in porphyry Cu deposits. Here we report the world's first occurrence of mantle-derived magmatic anhydrite in ultramafic rock from a magmatic platinum group elements sulfide deposit. The formation age of the clinopyroxenite and hornblendite hosting this anhydrite is 408.8 ± 2.1 Ma, shortly after the Paleozoic Oxygenation Event (POE). The anhydrite coexists with igneous carbonate minerals and Ca-O-C-Fe-S isotopes indicate oxidation of the mantle source by recycled oxidized surface-derived carbonates due to the POE. After the POE, the more oxidized supra-subduction mantle is coupled with the emergence of magmatic sulfide deposits in the orogenic belt during the Late Paleozoic (410–270 Ma) and a rapidly increased frequency of porphyry Cu deposits worldwide. Our results further suggest that high-Mg basaltic magma, typically considered to be the parent magma of magmatic sulfide deposits in orogenic belt, can evolve into PGE-enriched porphyry Cu deposit systems. This requires sufficiently high oxygen fugacity in magma after POE to retain sulfur predominantly as sulfate and thereby suppress sulfide saturation after olivine differentiation. Our findings link atmospheric oxygenation to deep Earth redox evolution and suggest that secular changes in Earth's redox state fundamentally influenced the evolution of key strategic metal sulfide deposits.

# Introduction

The Great Oxidation Event (GOE), mantle oxidation, and mega-mineralization events are coupled[1,2]. Compared with the GOE, during the Paleozoic Oxygenation Event (POE), the coupling relationship among atmospheric oxidation, mantle evolution, and important mineralization processes has not been well constrained. The atmospheric $O_2$ rose to near-modern levels after the POE (~410 Ma)[3,4]. Some evidence shows that the mantle was becoming more oxidized during this period[5,6]. Coincidentally, the world's largest magmatic sulfide nickel deposit in orogenic belt formed at 406.1 ± 2.7 Ma[7], and the abundance of porphyry copper deposits (PCD) began to increase rapidly after the POE[6,8]. Some PCD have high platinum group element (PGE) content[9], which are typically enriched in mantle rocks. The oxidized supra-subduction mantle is thought to play a key role in PCD[10]. This close association raises a fundamental question: what is the genetic link between atmospheric oxidation, oxidized supra-subduction mantle magma, and magmatic sulfide deposits and PGE-enriched PCD? Anhydrite, as a key link in Earth's Ca-O-S cycle, is an ideal mineral for revealing the above processes. Anhydrite inclusions exist in the upper mantle[11] and in the ultra-deep diamonds[12], especially in the supra-subduction mantle[11,13,14]. At the shallow upper crust, sedimentary anhydrite is a typical indicator mineral for atmospheric oxidation, and its addition to the magma promoted the formation of Bushveld, the world's largest PGE deposit, and Noril'sk, the world's largest magmatic nickel deposit[15,16]; while magmatic anhydrite in intrusions has been documented almost exclusively in intermediate to acidic rocks from PCD[17]. However, mantle-derived magmatic anhydrite (MMA) occurrence in ultramafic rocks has remained elusive. This gap is consequential: ultramafic rocks represent the closest approximation to primitive mantle melts, and the absence of MMA in such lithologies has precluded direct constraints on the oxidation state of sulfur in the supra-subduction mantle and its consequences for magmatic sulfide saturation and PGE behaviour during differentiation.

Here we report the world's first discovered MMA in ultramafic rocks. It was found in the 406 ± 3 Ma PGE-mineralized Langmuri ultramafic complex at the Eastern Kunlun orogenic belt (EKOB), Tibetan Plateau. The MMA occurs as disseminated grains in clinopyroxenite and hornblendite. Its sulfur isotope composition ($\delta^{34}S_{V\text{-}CDT}$ = 4.2‰ on average) and oxygen isotope signature ($\delta^{18}O_{V\text{-}SMOW}$ = 7.6‰ on average) record a mantle source. Based on the $\delta^{34}S_{V\text{-}CDT}$, $\delta^{18}O_{V\text{-}SMOW}$, $\delta^{13}C_{V\text{-}PDB}$, $\delta^{56}Fe_{IRMM014}$, $\delta^{44/44}Ca_{915a}$, and $\Delta'^{17}O$ values of whole-rock or minerals and the geological fact that anhydrite coexists with igneous calcite, it is inferred that the mantle source was oxidized by recycled surface-derived carbonates shortly after POE. This process is likely a key reason why the world's largest magmatic sulfide nickel deposits in orogenic belt and the world's first appearance of MMA in ultramafic rocks occurred during this period within the EKOB. We systematically compared the Pd-Ni characteristics of orogenic magmatic sulfide deposits and PGE-enriched PCD. Simulation calculations show that high-Mg basaltic magmas, typically considered to be the parent magma of magmatic sulfide deposits in orogenic belts, can evolve into PGE-enriched Ni-barren PCD systems. This requires sufficiently high oxygen fugacity to retain sulfur predominantly as sulfate and thereby suppress sulfide saturation after olivine differentiation. This study establishes the coupling relationship among POE, mantle oxidation, and key strategic metal sulfide deposits formation, providing new theoretical insights for Earth system science.

## MMA in ultramafic rocks from the Tibetan Plateau

Our research focuses mainly on the Langmuri PGE deposit, which is situated in the EKOB of the Tibetan Plateau (Extended Data Fig. 1a). The rock types in the Langmuri PGE deposit include dunite, phlogopite wehrlite (401.5 ± 8.1 Ma), phlogopite clinopyroxenite, anhydrite-bearing clinopyroxenite, hornblendite (408.8 ± 2.1 Ma), hornblende gabbro (414.9 ± 3.6 Ma) (Extended Data Figs. 1–6), and adakitic granite (408.8 ± 2.1 Ma)[18]. The intrusion order determined in this study is: dunite was earlier than phlogopite wehrlite and phlogopite clinopyroxenite, where the latter two lithofacies are in a compositional gradient relationship (Extended Data Fig. 2b and 2d); hornblende gabbro was obviously earlier than anhydrite-bearing clinopyroxenite and hornblendite (Fig. 1e), where the clinopyroxene gradually transforms into hornblendite. The EKOB area transferred from the oceanic crust subduction stage to the continental subduction period at about 438 Ma, and entered the post-collision extension stage around 428 Ma[19,20]. Therefore, the Langmuri complex formed in the post-collision stage.

This study discovered anhydrite in the ultramafic lithofacies from Drillcore DH0101 in the Langmuri PGE deposit (Fig. 1). The ultramafic lithofacies of DH0101 is divided into two sections, the upper ultramafic rock section from 65.5–350.7 m and the lower ultramafic rock section from 478–717 m, respectively. Among them, 595–607 m is anhydrite-bearing clinopyroxenite, and the rest is hornblendite (Fig. 1a). Anhydrite exists from 129–141 m and 402.6–727 m. It is disseminated distributed in clinopyroxenite (Fig. 1i) and hornblendite (Fig. 1d and 1f), existing in the intergranular spaces of clinopyroxene or amphibole (Fig. 1g and 1j), coexisted with magnetite (Fig. 1g) or pyrite (Fig. 1j), or occasionally as inclusions in pyrite or associated with calcite (Fig. 1h). For the lower ultramafic rock section, the $Mg^{\#}$ of the amphibole is highest in the middle and decreases towards the top and bottom (Fig. 2, Supplementary Table 2). Combined with the presence of clinopyroxenite in the middle of the lithofacies, it is inferred that the ultramafic rocks crystallize first in the middle and later at the top and bottom. The same pattern applies to the upper ultramafic rock section. Calculations by an oxygen fugacity meter for calcium amphibole[21] show that the oxygen fugacity of the magmatic system gradually decreases during crystallization differentiation. For the lower ultramafic rock section, the oxygen fugacity (in $\lg fo_2$) gradually decreases from -6.0 in the middle to -8.7 at the top and to -7.9 at the bottom (Fig. 2, Supplementary Table 3). This is also supported by the iron isotopes of pyrite and magnetite. Previous studies have shown that redox processes lead to the fractionation of Fe isotopes, with ferric iron showing heavy isotope enrichment relative to ferrous iron[22]. The lower ultramafic rock section shows that the $\delta^{56}Fe_{IRMM014}$ of pyrite and magnetite are highest at the early crystallization sites, gradually decreasing towards the top and bottom (Fig. 2), indicating that the oxygen fugacity decreases during crystallization. The $\delta^{34}S_{V\text{-}CDT}$ value of anhydrite decreases from the center part to the marginal part, dropping from 9.5‰–12.4‰ to a minimum of -12.1‰. Similarly, the $\delta^{34}S_{V\text{-}CDT}$ value of pyrite shows a similar trend in the lower ultramafic rock section, although it is less pronounced in the upper ultramafic rock section (Fig. 2). The mean $\delta^{34}S_{V\text{-}CDT}$ of anhydrite and pyrite is 4.2‰ (n = 25) and -3.4‰ (n = 49), respectively (Supplementary Table 4). The mean $S^{6+}$ and $S^{2-}$ content in the rock is 2.29 wt.% and 1.45 wt.% (n = 7), respectively (Supplementary Table 5). Thus, it can be inferred that the $\delta^{34}S_{V\text{-}CDT}$ of the magma is about 1.3‰. This undoubtedly indicates that the S in the anhydrite in this study originated from the mantle, which is completely different from the sulfur isotope characteristics of Cambrian-Early Devonian marine sulfates ($\delta^{34}S_{V\text{-}CDT}$ >20‰)[23]. The average $\delta^{18}O_{V\text{-}SMOW}$ values of ocean sulfates from the Late Proterozoic, Cambrian, and Devonian are 17.2‰, 13.9‰, and 15.4‰, respectively[23]. While the $\delta^{18}O_{V\text{-}SMOW}$ values of the anhydrites in Langmuri are distributed between 7.4‰ and 7.7‰, which is completely different from that of sedimentary sulfates and is a typical oxygen isotope characteristic of magmatic anhydrite (Supplementary Table 1).

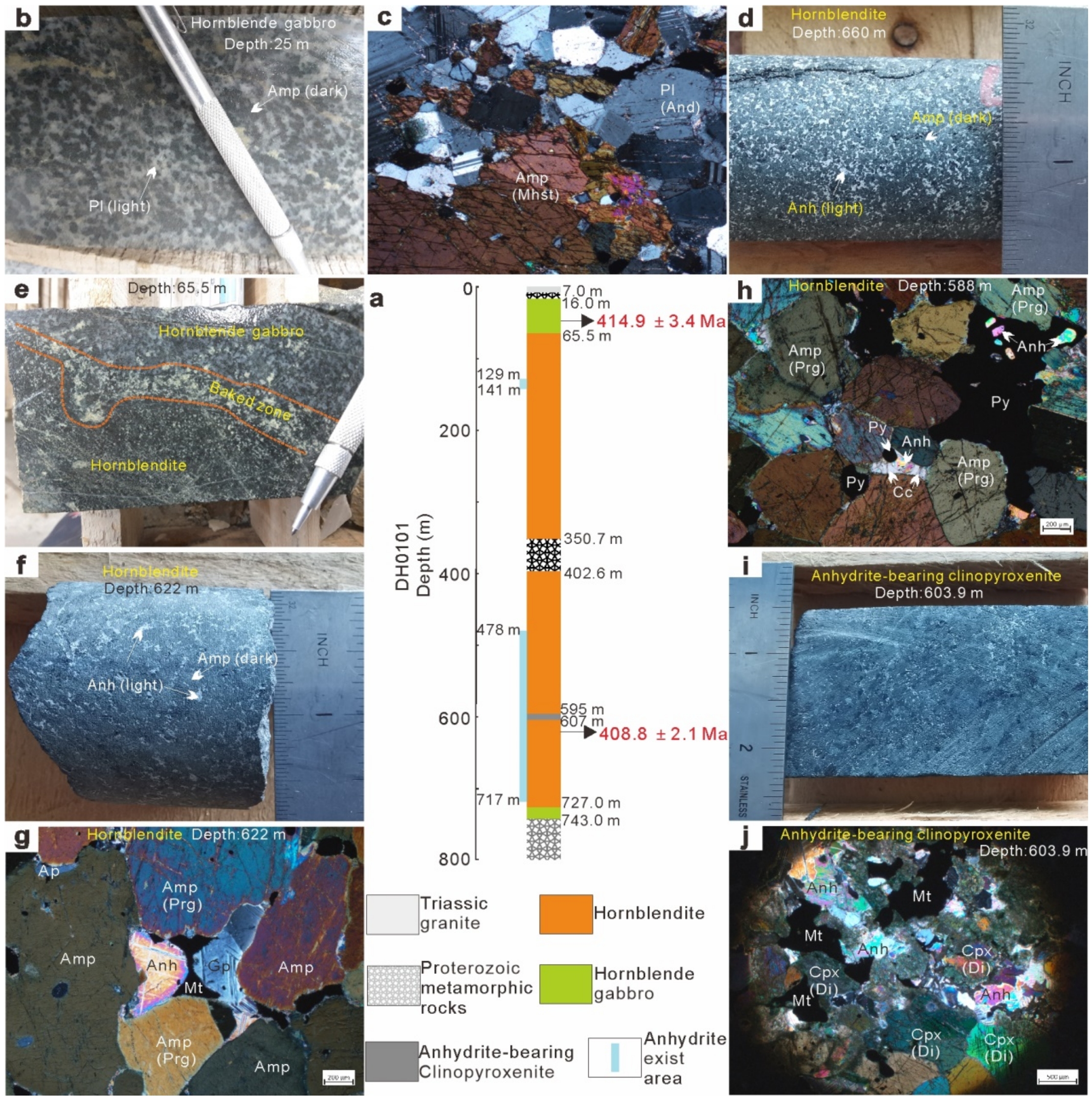

**Figure 1. Lithofacies Distribution and Rock Characteristics of drill hole DH0101 in Langmuri PGE deposit.** **a,** Lithofacies distribution of borehole DH0101. **b,** Hand specimen of hornblende gabbro. **c,** Microscopic feature of hornblende gabbro. **d,** Hand specimen of gypsum-bearing hornblendite (depth 660m). **e,** The hornblendite baked the hornblende gabbro, indicating the hornblende gabbro was earlier than the hornblendite. **f**, Gypsum-bearing hornblendite (depth 622m). **g,** Anhydrite coexists with gypsum and magnetite in hornblendite; **h,** Anhydrite was associated with calcite or encased in pyrite.

**i,** Hand specimen of gypsum-bearing clinopyroxene.

**j,** Microscopic characteristics of gypsum-bearing clinopyroxene, and its pyroxene type is diopside. Amp = amphibole, Mhst = Magnesio-hastingsite, Prg = Pargasite, Cpx = clinopyroxene, Di = diopside, Pl = plagioclase, And = andesite, Anh = anhydrite, Gy = gypsum, Mt = magnetite, and Cc = calcite.

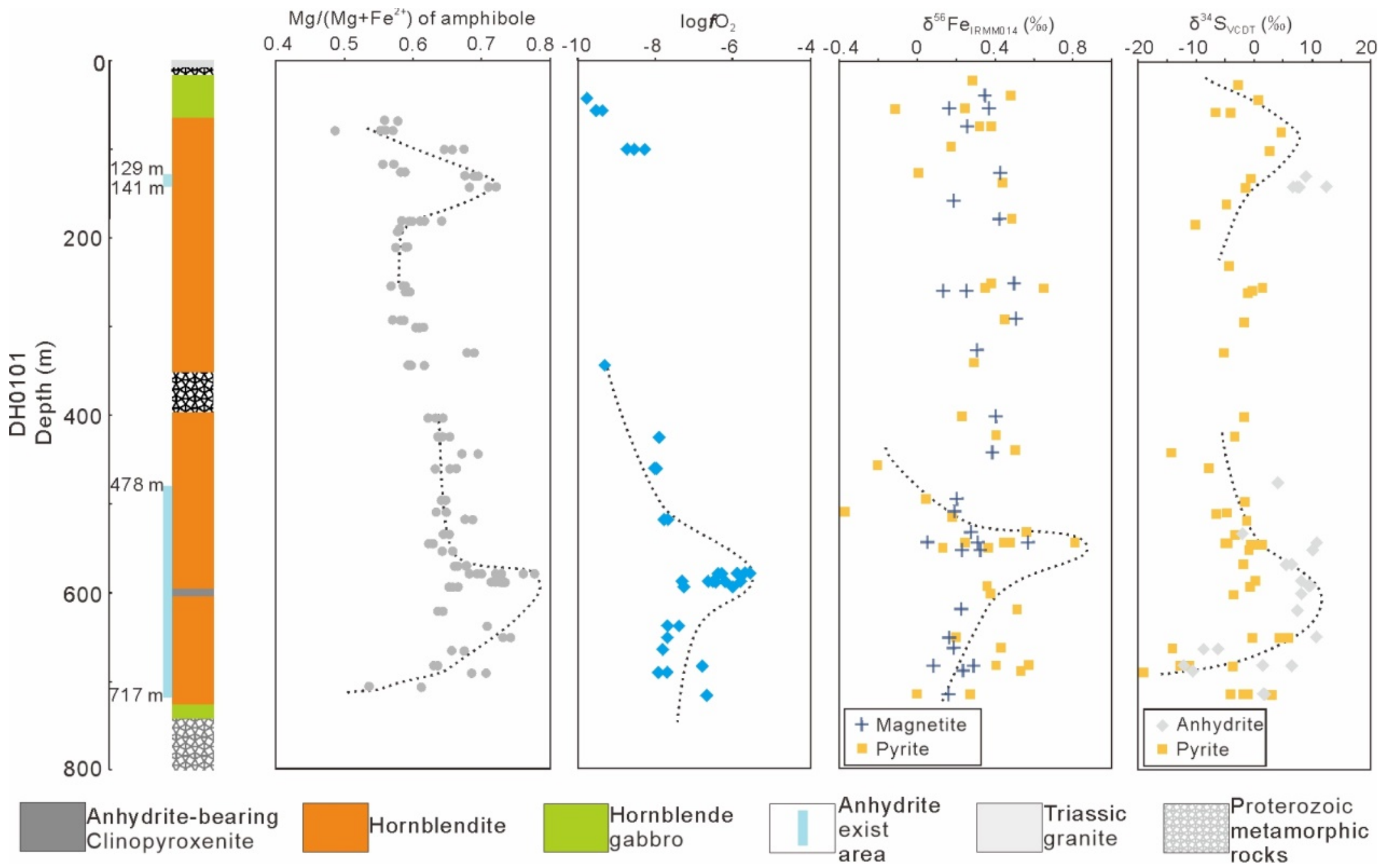

**Figure 2. Spatial variation of geochemical composition in DH0101.**

The $\delta^{18}O_{V\text{-}SMOW}$ values of the phlogopite wehrlite, phlogopite clinopyroxene, and hornblendite in the Langmuri deposit range from 9.7 to 12.3‰, 10.5‰ to 13.0‰, and 6.6‰ to 10.6‰, respectively (Extended Data Fig. 7a, Supplementary Table 4). Their $\delta^{13}C_{V\text{-}PDB}$ values range from -2.9 to -6.2‰, -4.0 to -7.3‰, and -3.0 to -5.0‰, respectively. On the $\delta^{18}O_{V\text{-}SMOW}$ *vs.* $\delta^{13}C_{V\text{-}PDB}$ diagram, these rocks are located between the upper mantle and sedimentary carbonate rocks (Extended Data Fig. 7a), indicating that sedimentary carbonate rock components have been incorporated into the Langmuri complex. In the hornblendite, calcite occasionally coexists with anhydrite (Extended Data Fig. 8a), or encased in amphibole (Extended Data Fig. 8b), or appears alone in the interstices of amphibole (Extended Data Fig. 8c). In the Xiarihamu Ni deposit in the EKOB, the largest Ni deposit in the orogenic belt, clinopyroxene encases calcite and dolomite in dunite (Extended Data Fig. 8d). The average Mn content of calcite in the ultramafic rocks from Langmuri and Xiarihamu deposits is 2764 ug/g and 1162 ug/g, respectively, and the Sr content is 1598 ug/g and 2465 ug/g, respectively (Supplementary Table 2). This is consistent with the Mn and Sr content characteristics of igneous carbonate minerals, Mn ≥ 1161 ug/g and Sr ≥ 1268 ug/g[24]. The chondrite-normalized distribution pattern of the above carbonate minerals is significantly different from that of the surrounding Proterozoic marble (Extended Data Fig.9a). The δEu of the surrounding Proterozoic marble is less than 0.59, while the δEu of carbonate minerals in the ultramafic rocks from Langmuri and Xiarihamu deposit is greater than 0.6, with mean values of 0.84 and 2.82, respectively (Supplementary Table 2). Except for one calcite grain, which coexists with anhydrite and has a $[La/Yb]_N$ ratio of less than 1, the $[La/Yb]_N$ ratio of the other calcites in ultramafic rocks ranges from 10 to 95, which is much higher than the $[La/Yb]_N$ ratio (6–7) of the surrounding Proterozoic marble. The $[La/Yb]_N$ ratio of MMA ranges from 364 to 414, and it is highly enriched in light rare earth elements (Supplementary Table 2, Extended Data Fig. 9c-d), which may explain the low $[La/Yb]_N$ ratio in calcite associated with anhydrite. Thus, this paper hypothesizes that the C-O isotopic characteristics of the ultramafic rocks are not due to contamination by the surrounding marble, but rather to the metasomatism by carbonatite melt in the source region.

The $\delta^{44/42}Ca_{915a}$ values of phlogopite wehrlite and phlogopite clinopyroxenite are 0.58-0.66‰ and 0.65-0.69‰, respectively (Extended Data Fig. 7b, Supplementary Table 4), which are higher than the values of the upper mantle (0.46‰ ± 0.02%)[25]. While, the $\delta^{44/42}Ca_{915a}$ values of MMA in Langmuri are -0.27‰ to -0.31‰, which are lower than those of the surrounding Proterozoic schist (0.34‰) and marble (0.34‰), as well as those of typical intraplate carbonate rocks (0.04‰–0.16‰), post-collision carbonate rocks (-0.03‰ to 0.23‰), and modern ocean island carbonate rocks (0.15‰ to 0.19‰) worldwide[26]. It is quite different from the crust contamination of upper mantle magma trend (Extended Data Fig. 7b). A reasonable explanation for this feature is isotope Rayleigh fractionation during fractional crystallization. The overall trend of heavy calcium isotope enrichment capacity is: olivine and clinopyroxene > calcic amphiboles > anhydrite [27,28]. This result in olivine and clinopyroxene formed at early crystallization stage have higher $\delta^{44/42}Ca_{915a}$ value than the upper mantle, while MMA formed during last stage exhibits a much lower $\delta^{44/42}Ca_{915a}$ value (Extended Data Fig. 7b). The $\delta^{18}O_{V\text{-}SMOW}$ values of amphibole and MMA are positively correlated with $\delta^{44/42}Ca_{915a}$, which is largely consistent with the correlations of intraplate carbonatites, post-collisional extensional carbonatites, and fresh carbonatites in the East African Rift (Extended Data Fig.7b) rather than the negative correlation for the crustal contamination process. This perhaps also reflect the source region has carbonatite melt metasomatism.

# Coupling of the POE, mantle Oxidation, and magmatic sulfide deposits in orogenic belts

During the GOE, the oxygen fugacity of the sub-arc mantle changed due to sediment recycling, clearly marked by a shift in zircon εHf(t) value from positive to negative or zero values in tonalite-trondhjemite-granodiorite rocks before and after 2.3 Ga[2]. The same feature exists in zircon εHf(t) values of mafic-ultramafic rocks before and after the POE (410 Ma) in the EKOB (Extended Data Fig. 10, Supplementary Table 6), indicating that the mantle wedge became more oxidized. The $Fe^{3+}/\Sigma Fe$ in basaltic melts of different ages globally revealed a jump in mantle oxygen fugacity during the Early Devonian (Fig. 2c). All evidence points to the coupling of atmospheric oxidation and mantle oxidation. There is controversy regarding whether atmospheric oxidation or mantle oxidation is the cause or the result. Some theories suggest that deep Earth processes lead to atmospheric oxygenation[29], while others argue that atmospheric oxidation is the cause and deep Earth oxidation is the effect; Atmospheric oxidation leads to more oxidizing ions in sedimentary rocks, and the sedimentary rocks subduction leads to mantle oxidation[30]. In this study, the Δ'17O of MMA in the Hornblendite (408.8 ± 2.1 Ma) ranged from -0.13‰ to 0.01‰ (Fig. 3b, Supplementary Table 4). The Δ'17O values in sedimentary rocks older than 410 Ma is significantly lower than -0.2‰[4]. If the mantle wedge was oxidized before 410 Ma and then led to atmospheric oxidation, according to the mechanism that subduction causes mantle oxidation[31], the mafic-ultramafic rocks should bear the imprint of crustal material subducted before 410 Ma. That is, MMA should have a $\Delta'^{17}O$ anomaly below -0.2‰. However, the results in this study do not support this hypothesis. Therefore, this paper infers that atmospheric oxidation likely preceded mantle oxidation during the POE. Combining the evidence showing carbonatite melt metasomatism in the source region and the signal of sedimentary carbonate on the $\delta^{18}O_{V\text{-}SMOW}$ *vs.* $\delta^{13}C_{V\text{-}PDB}$ diagram, the most reasonable explanation probably is as follows. After 410 Ma, although the EKOB was in a post-collision extensional stage, subduction did not completely cease. Due to atmospheric oxidation, sedimentary carbonates carried more oxidizing ions into the mantle during subduction, leading to the oxidation of the super-subducting mantle and the formation of anhydrite-bearing ultramafic rocks at 408.8 ± 2.1 Ma. The transition from atmospheric oxidation to mantle oxidation in subduction zones happen within only 1–2 Ma[31].

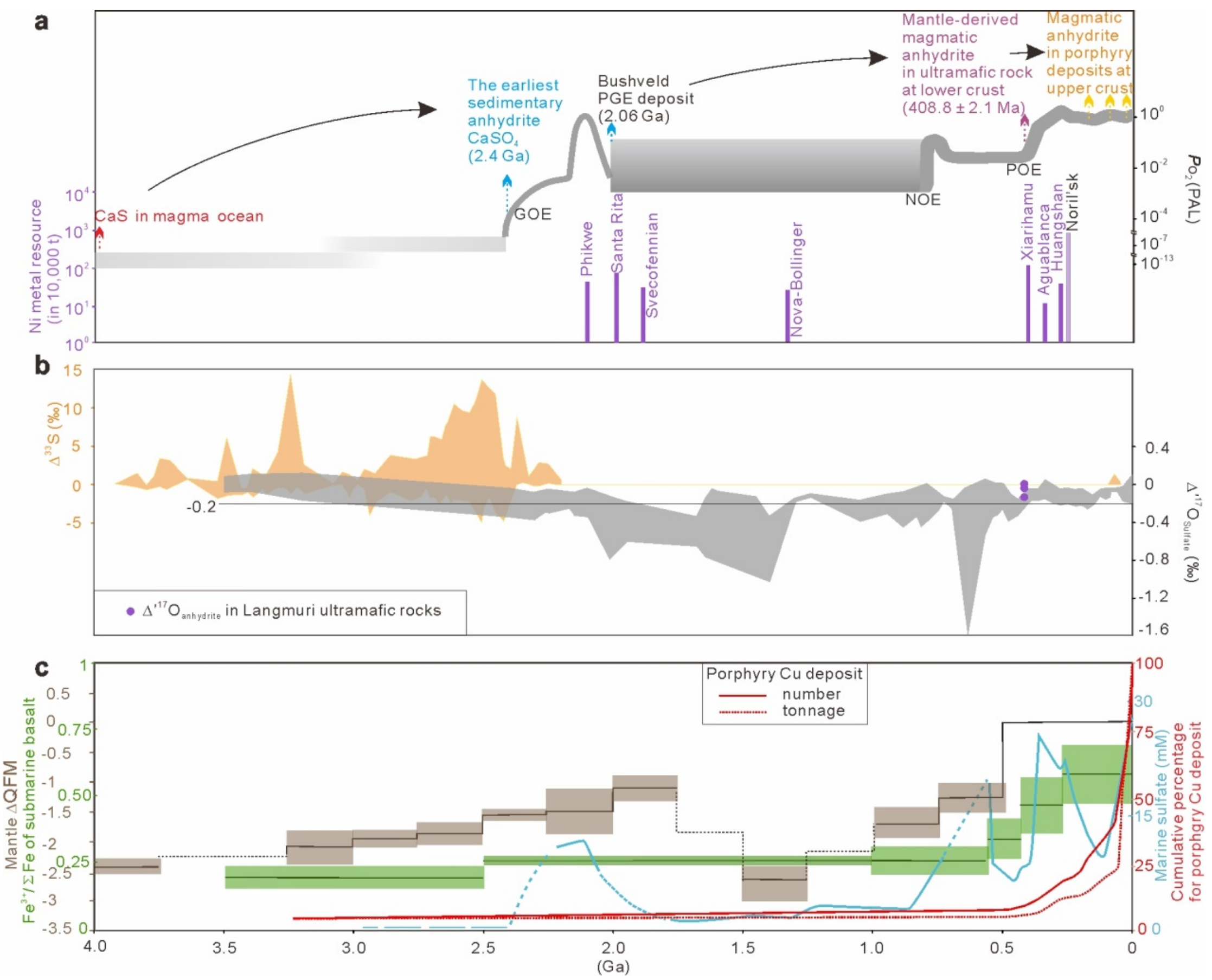

**Figure 3. The relationship between the mantle and atmospheric oxidation and typical sulfide ore deposits.**

**a,** CaS in magma ocean (ref 32); The world's earliest sedimentary anhydrite was found in the upper Duitschland Formation 2.4 Ga (ref. 33); Most magmatic anhydrite in porphyry Cu deposits are later than 168 Ma (ref. 17); The dark purple pillars represent the top seven magmatic nickel deposits in the orogenic belt, while the light purple pillar represent the world's largest nickel deposits, Noril'sk, which is of mantle plume origin (Age, reserve, and geological setting references are listed in Supplementary Table 7). GOE time is from refs.30,34, NOE and POE time are from ref. 35 and ref. 4, respectively.

**b,** Integrated sedimentary$\Delta^{33}S$ and $\Delta'^{17}O_{sulfate}$ records are from ref. 4 and references therein. The black line represents the −0.2‰ $\Delta'^{17}O_{sulfate}$ values, the lower limit observed in sulfate after POE (ref 4).

**c,** Temporal correlation between global porphyry Cu deposits and mantle-ocean-atmospheric evolution through geological time. Mantle oxygen fugacity is from ref. 5; The $Fe^{3+}/\sum Fe$ ratios (in green) of the submarine basalts are from ref. 6; The temporal distribution of global porphyry Cu deposits based on the database from ref. 8; Marine sulfate contents on average (in blue; uncertainty not marked for simplification) are from refs. 36,37; the blue dashed line represents no data being reported.

There are four significant sulfur cycle events globally: CaS existed in the magma ocean before GOE[32]; sedimentary anhydrite (2.4 Ga) began to appear on Earth's surface during or after GOE[33]; MMA in ultramafic rocks formed at 408 Ma shortly after GOE; and the widespread magmatic anhydrite in the upper crustal PCD after 168 Ma[17]. These four landmark events in the sulfur cycle are closely related to atmospheric and mantle oxidation and have also influenced the mineralization processes of key strategic metals. The top seven Ni deposits in the orogenic belt are concentrated in the Paleoproterozoic and Early Devonian-Early Permian periods (Fig. 3a). Among them, the second largest, Santa Rita, the third largest, Phikwe, and the fifth largest, Svecofennian, formed at 1.99 Ga, 2.0–2.6 Ga, and 1.88 Ga, respectively (Supplementary Table 7), during or slightly after the GOE (Fig. 3a). The first largest, Xiarihamu, the fourth largest, Huangshan, and the seventh largest, Aguablance, formed at 406.1 ± 2.7, 282 ± 20, and 336.2 ± 1.7 Ma, respectively (Supplementary Table 7), slightly later than the POE, during the period of rising atmospheric oxygen content from 410 to 270 Ma (Fig. 3a). These two periods also correspond to the stage of rapid mantle oxidation (Fig. 3c). Diamond inclusions research and high *P*-*T* experiments have found that the carbonatite melt metasomatism in mantle can promote the release of metals such as nickel from mantle peridotite into the melt[38,39].

Combined with the geological fact that igneous carbonatite minerals exist in the Xiarihamu Ni deposit and the Langmuri PGE deposit (Extended Data Fig. 8), it is reasonable to speculate that the carbonatite melt metasomatism may be a key step in the mineralization process of magmatic sulfide deposits in orogenic belt during or after the POE.

# The intrinsic connection between magmatic PGE sulfide deposits and PGE-enriched porphyry Cu deposits

The vast majority of magmatic anhydrite in intrusions is found in PCD[17] (Supplementary Table 1). The platinum group minerals in the PCD are predominantly Pd, Te, and Bi compounds[9], which is nearly the same to the platinum group mineral types in the Langmuri PGE sulfide deposit (Extended Data Figs. 11–12, Supplementary Table 2). Moreover, adakitic granites, important indicative magmas for PCD[40,41], exist in Langmuri deposit contemporaneous with the ultramafic complex[18]. There is a homologous crystallization evolution trend from ultramafic rocks, mafic rocks, and adakitic granites (Extended Data Fig. 13). These prompted us to investigate the intrinsic connection between magmatic PGE sulfide deposits and PGE-enriched PCD in orogenic belts.

Based on the whole-rock Pd content in PCD, we divided them into PGE-enriched PCD and PGE-barren PCD, using Pd = 10 ppb as the boundary (Fig. 4a). The whole-rock Pd contend and Pd content in 100% sulfides of PGE-enriched PCD are comparable to or slightly higher than the Pd content of magmatic PGE deposit in orogenic belt; the Pd content in 100% sulfides of PGE-barren PCD is comparable to that in magmatic PGE deposit in orogenic belt, ranging from 5 to 100 ppb (Fig. 4a, Supplementary Tables 8–10). Figure 4a shows a positive correlation between Pd and Ni content in both orogenic magmatic PGE sulfide and Ni deposits. This paper estimates that the parent magma of the Langmuri platinum group metal deposit is a high-Mg basaltic magma (MgO = 11.49 wt.%, FeO = 7.54 wt.%, $SiO_2$ = 51.96 wt%, Extended Data Fig. 14). Using high-Mg basaltic magma containing 14 ppb Pd and 300 ppm Ni as the parent magma, segregation calculations show that the Pd and Ni contents of 100% sulfides in orogenic magmatic PGE deposits are distributed within the range from no sulfide segregation to those that have undergone 0.002% sulfide segregation (Fig. 4b). In contrast, the Pd and Ni contents of 100% sulfides in orogenic magmatic Ni deposits are located within 0.002%–0.005% sulfide segregation (Fig. 4b, Supplementary Table 11).

Given that the partition coefficient of Pd between sulfide and silicate melts ranges from $6.7 \times 10^4$ to $5.36 \times 10^5$ [42], sulfide segregation would significantly reduce the Pd content in the magma. Therefore, it can be considered that PGE-enriched PCD could not have evolved from high-Mg basaltic magma after sulfide segregation; and if the degree of sulfide segregation during the evolution of high-Mg basaltic magma is >0.002%, the resulting magma is unlikely to become the parent magma of PGE-barren PCD. PGE-barren PCD may have received metallic elements from magma with low-degree sulfide segregation (less than 0.002%) (Fig. 4b). The whole-rock Ni content and Ni contents in 100% sulfides of PCD are about two orders of magnitude lower than those in magmatic deposits in orogenic belt (Fig. 4a and 4b). In contrast, although both Ni and Pd are chalcophile elements[42,43], Pd content in PCD is on the same order of magnitude as in magmatic deposits in oregenic belt. This decoupling of Ni and Pd elements is probably due to the parent magma of PCD was the evolution result of the high-Mg basaltic magma that underwent olivine fractional crystallization with no sulfide saturation. Simulations show that high-Mg basaltic magma containing 14 ppb Pd and 300 ppm Ni can evolve into Magma B with 22.98 ppb Pd and 1.81 ppm Ni after 40% olivine fractional crystallization under no sulfide saturation condition (Fig. 4c, Supplementary Table 12). The Ni and Pd contents of Magma B after 0-0.005% sulfide segregation are consistent with the Ni and Pd contents of most PGE-enriched PCD and some PGE-barren PCD (Fig.4b). That is, the parent magma of magmatic sulfide deposits in orogenic belts can evolve into the parent magma of PCD. The Dy/Yb ratio of Langmuri samples evolved towards PCD region with increasing $SiO_2$ also supports the above opinion (Fig. 4d). Generally, after olivine fractional crystallization in high-Mg basaltic magma, the temperature drops significantly and the FeO content of the magma system is consumed, reducing the minimum sulfur content required for magma sulfide saturation to below 1000 ppm or 700 ppm[20,44], which is easy for the sulfide saturation achievement. A key factor to prevent sulfide saturation may be that the oxygen fugacity of the magma system is greater than the upper limit of sulfide existence ($S^{2-}$–$S^{6+}$ equilibrium oxygen fugacity), that is, S in the melt exists mainly as sulfate ions. On the $SiO_2$-Dy/Yb plot, samples from the Langmuri deposit show a faster Dy/Yb decreasing trend than amphibole crystallization differentiation alone (Fig. 4d). The average Dy/Yb ratios of anhydrite, amphibole, apatite, and clinopyroxene in the Langmuri deposit are 10.3 (n = 6), 2.7 (n = 6), 3.5 (n = 6), and 2.3 (n = 4), respectively (Supplementary Table 2). This indicates that anhydrite crystallization differentiation can rapidly reduce the Dy/Yb ratio of the magmatic system. Thus, the trend on the $SiO_2$-Dy/Yb plot of Langmuri samples is dominated by the co-crystallization of amphibole and anhydrite (Fig. 4d). The Langmuri complex formed during the initial stage of increased atmospheric oxygen content and higher mantle oxidation between 410 and 270 Ma, experienced a small amount of sulfide saturation after mineral crystallization differentiation, resulting in disseminated sulfide ore. With the increase of mantle oxygen fugacity after POE (Fig. 3c), especially supra-subduction mantle oxygen fugacity, high-Mg basaltic magmas, even after olivine and clinopyroxene crystallization differentiation, may not have experienced sulfide saturation due to the higher oxygen fugacity of the magma system. This allows PGE and Cu to accumulate in melt, ultimately forming PGE-enriched PCD. This may also be one of the reasons why the frequency of PCD increased rapidly after the POE (Fig. 3c). In addition, this model can also explain why the Pd/Pt ratio of most PCD is significantly greater than 1[9]. Under high oxygen fugacity, without sulfide saturation, Pt can combine with As and other elements to form alloys[45], consuming Pt abundance. The Langmuri ultramafic rocks, serving as a relay station for the transformation of mantle-derived magma into upper crustal rocks, show $PtAs_2$ encapsulated in clinopyroxene (Extended Data Fig. 15), indicating that $PtAs_2$ crystallized earlier than or co-crystallized with clinopyroxene. Besides, Pt metal enclosed in clinopyroxene is even visible in Langmuri deposit (Extended Data Fig. 16). Meanwhile, sulfide in the Langmuri deposit occurred later than olivine and pyroxene, and is roughly co-crystallized with phlogopite (Extended Data Fig. 2d), undoubtedly demonstrating that Pt minerals can accumulate even without sulfide saturation. In summary, the opinion that high-Mg basaltic magma can evolve into the parent magma of PCD can well explain the Ni-Pd-Pt characteristics in PCD.

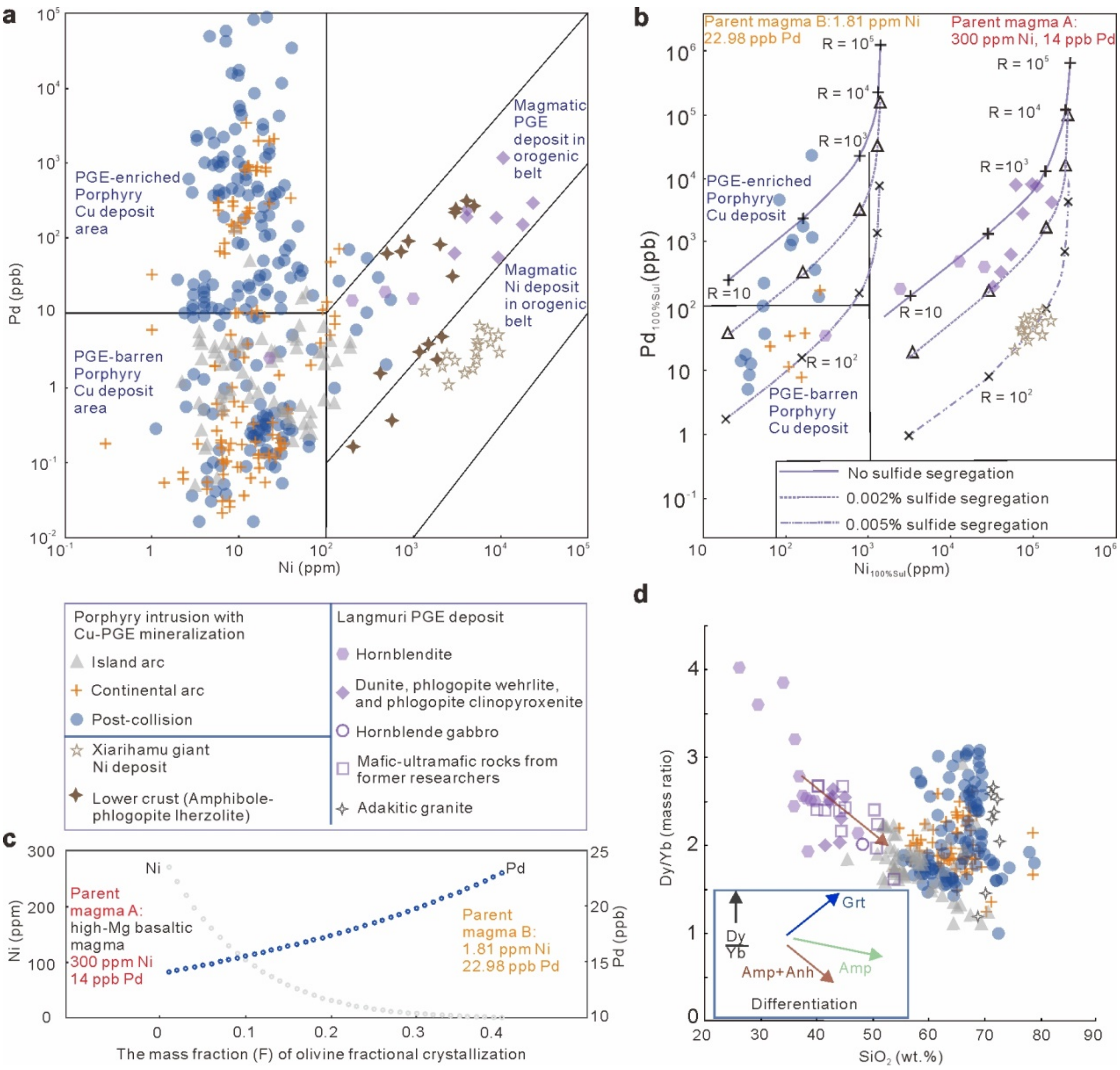

**Figure 4. The intrinsic connection between magmatic PGE sulfide deposit and PGE-rich PCD.**

**a,** Whole-rock Ni-Pd elemental comparison between PCD and the Langmuri magmatic PGE sulfide deposit, lower crust peridotite, and the largest magmatic Ni sulfide deposit in orogenic belt (Xiarimamu magmatic Ni deposit).

**b,** Model calculations of the variation in $Pd_{100\%Sul}$ with $Ni_{100\%Sul}$ of sulfides from magmatic sulfide deposits and PCD. The Pd and Ni concentrations of the sulfides are corrected to 100% sulfide content (Methods, Supplementary Tables 9–10). The right solid line represents the sulfide melts segregated from PGE-rich magma [Parent magma A (high-Mg basaltic magma)] with 14 ppb Pd and 300 ppm Ni under variable R values (labeled in the diagram). The left solid line displays the sulfide liquids separated from parent magma B with 22.98 ppb Pd and 1.81 ppm Ni (Methods, Supplementary Table 11). Parent magma B is the product of parent magma A after 40 wt% olivine crystallization. The dashed line and dashed-dotted line show the sulfide liquids produced by sulfide immiscibility in Pd-depleted magma due to 0.002% and 0.005% sulfide removal from Pd-rich magma, respectively.

**c,** The changing trends of Ni and Pd content in the residual magma during the olivine fractionation crystallization process of high-Mg basaltic magma (parent magma A) (Methods, Supplementary Table 12).

**d,** Dy/Yb *vs.* $SiO_2$. The crystallization differentiation trend of garnet and amphibole is from [46].

Grt = garnet, Amp = amphibole, and Anh = anhydrite.

Data source: Lower crust peridotite is from ref. [10]; other original data and source are listed in Supplementary Table 5, 8–10.

# Extremely high Sr/Y and La/Yb ratios in MMA challenge classical adakitic rock formation theory

Both subducted oceanic crust type (O-type) and thickened lower crust type (C-type) adakitic rocks are considered to have source regions greater than 1 GPa and contain garnet as a residual phase[47-49]. This view is based on the fact that the partition coefficients of heavy rare earth elements and Y between garnet and coexist melt are significantly greater than 1, while the partition coefficients of light rare earth elements and Sr in garnet and coexist melt are much less than 1[50,51]. This explains the high Sr/Y mass ratio > 40 and La/Yb mass ratio > 20 ratios in aidakitic rocks. The Sr/Y and La/Yb mass ratios of MMA in Langmuri deposit range from 895 to 2840 and 507 to 1871, respectively, with average values of 1690 and 965 (Supplementary Table 2). The formation of intermediate-acidic magmas with high Sr/Y ratios > 40 and La/Yb ratios > 20 could be achieved simply by the MMA entering the intermediate-acidic magma as a melt or fluid during partial melting or metasomatism. Anhydrite-bearing intrusive rocks have high Sr/Y ratios in PCD[17], which is consistent with the fact that MMA has extremely high La/Yb and Sr/Y ratios. Garnet as a residual phase in the source region of adikitic rocks may not be a must.

# Methods

## Whole-rock major-element analysis

All samples analysed in this study were collected from drill holes. Major-element concentrations were determined by wavelength-dispersive X-ray fluorescence (WD-XRF) spectrometry using a PANalytical Axios 4.0 kW instrument, and trace-element concentrations were measured by inductively coupled plasma mass spectrometry (ICP-MS), following Chinese national standards GB/T 14506.28-2010, GB/T 14506.2-2010, GB/T 14506.30-2010 and GB/T 14506.14-2010. The majority of samples were analysed at the Xi'an Center of China Geological Survey. Major and trace elements of seven hornblendite samples from drill hole DH0101 in the Langmuri deposit and Proterozoic marble samples from the Xiarihamu deposit were analysed at ALS Chemex (Guangzhou) Co. Ltd. Whole-rock sulphur contents were determined by a gravimetric method and infrared absorption (IR08) using an infrared sensor at 1,350 °C, with a lower detection limit of ~0.01 wt%. Sulphate sulphur ($S^{6+}$) was extracted by leaching with sodium carbonate solution, precipitated as barium sulphate and determined gravimetrically (method S-GRA06; detection limit ~0.01 wt%). Sulphide sulphur ($S^{2-}$) was analysed by selective carbonate extraction followed by infrared detection using a Leco analyser (method S-IR07; detection limit ~0.01 wt%). Whole-rock copper and nickel contents were measured by inductively coupled plasma atomic emission spectroscopy (ICP-AES). Analytical precision was ±8% for S and ±3% for Ni and Cu.

Inorganic carbon in marble samples was determined by perchloric acid ($HClO_4$) digestion followed by coulometric titration with ethanolamine colorimetric detection (method C-GAS05). Organic carbon was determined by acid washing to remove inorganic carbon, followed by filtration, drying, and infrared detection using a LECO analyser at 1,350 °C (method C-IR06). The results of major and trace elements are listed in Supplementary Table 5.

## PGE content analysis

Twelve samples were collected from drill core DH0101, DH2403, DH1201, and DH1703, and ZK1903, for PGE analyses. The location and the results have been listed in Supplementary Table 9. Those samples were crushed in steel jaws to −10 mesh, and then ~200 g of this fraction was ground to −200 mesh powder using an agate mortar. The PGE concentrations were determined at the Institute of Geochemistry, Chinese Academy of Sciences by isotope dilution (ID)-ICP-MS, using an improved digestion technique with beakers sealed in a stainless steel pressure bomb[52]. The total procedural blanks varied from 0.009 ng (Ir) to 0.033 ng (Pd), and the lower limits of detection ranged from 0.004 ppb (Ir) to 0.012 ppb (Pd)[52]. The precision and accuracy were monitored by analysis of the standards WGB-1 and UMT-1.

## Major element compositions of Minerals

The major-element compositions of most minerals were determined using a field-emission JXA-iHP200F electron probe microanalyzer (EPMA) at the Institute of Deep-sea Science and Engineering, Chinese Academy of Sciences. Partial amphibole analyses were conducted at Wuhan SampleSolution Analytical Technology Co., Ltd. (on a JXA-iHP200F EPMA). Carbonate-mineral analyses for the Xiarihamu carbonatite were performed at Chang'an University (using a three-spectrometer JXA-8100 EPMA), and selected analyses of olivine, clinopyroxene and metallic minerals were carried out at the Xi'an Centre of Geological Survey, China Geological Survey (on a JEOL JXA-8230 EPMA). Analytical conditions for anhydrite were 15 kV and 5 nA; for calcite, dolomite and apatite, 15 kV and 5–10 nA; and for olivine, clinopyroxene, amphibole and metallic minerals, 15 kV and 20 nA. The beam diameter ranged from 1 to 10 μm for most phases, with peak and background counting times of 20 s and 10 s, respectively. The calibration standard samples for the content of major elements use 53 kinds of mineral standard samples, 44 kinds of elemental standard samples and 15 kinds of rare earth element standard samples provided by SPI Company. The data correction method adopts the ZAF correction method of JEOL. The results of major elements of minerals are listed in Supplementary Table 2.

## In situ LA-ICP-MS trace elements analysis

Trace elements of amphibole, apatite, anhydrite, calcite, and dolomite were determined using a NWR 193 nm ArF Excimer laser-ablation system coupled to an iCAP RQ (ICPMS) at Guangzhou Tuoyan Analytical Technology Co., Ltd., Guangzhou, China. The ICPMS was tuned using NIST 610 standard glass to yield low oxide production rates. 0.7 l/min He carrier gas was fed into the cup, and the aerosol was subsequently mixed with 0.89 l/min Ar make-up gas. The laser fluence was 5.0 J/cm$^2$, with a repetition rate of 6 Hz, a 60 μm spot size and analysis time of 45 s, followed by a 40 s background measurement. Trace element compositions of samples are calibrated against various reference materials (NIST SRM 610 and BCR-2G) using a total normalization method[53]. The glass standard SRM 612 and BIR-1G are also detected to evaluate the data quality[54,55]. The raw isotope data were reduced using the 3D Trace Element data reduction scheme of IOLITE software[56]. The results of minor elements of minerals are listed in Supplementary Table 2.

## Zircon U-Pb dating

Three samples, including hornblende gabbro from drill hole D0101 at a depth of 50m, hornblendite from drill hole DH0101 at a depth of 620m, and phlogopite peridotite from drill hole DH2403 at a depth of 128–138m, were used for zircon U-Pb dating. Zircon grains were separated from bulk samples by conventional heavy-liquid and magnetic techniques and then purified by hand-picking under a binocular microscope. Representative zircon grains were embedded in epoxy resin and then polished to expose their internal structures. Zircon cathodoluminescence (CL) images of hornblendite were obtained at the Langfang Tuoxuan Testing Technology Co., Ltd., China, and CL images of hornblende gabbro and phlogopite peridotite were obtained at Guangzhou Tuoyan Testing Technology Co., Ltd. Zircon U-Pb dating was carried out by LA-ICP-MS at the Key Laboratory for the Study of Focused Magmatism and Giant Ore Deposits, Ministry of Natural Resources, using a Geolas 2005 excimer ArF laser-ablation system coupled with an Agilent 7700x ICP-MS[57]. The LA-ICP-MS zircon U-Pb dating data are listed in Supplementary Table 13.

## In-situ Hf isotope analyses

*In situ* Lu-Hf isotope analysis of zircon grains from hornblende gabbro and hornblendite was conducted at the Key Laboratory for the Study of Focused Magmatism and Giant Ore Deposits, Ministry of Natural Resources using a Nu-plasma multi-collector ICP–MS (MC–ICP–MS) coupled with a GeoLas 2005 193 nm ArF excimer laser. Sample sites for Lu-Hf isotope analyses were selected to coincide with areas that were analyzed during U-Pb dating. The spot size and laser pulse frequency were 44 μm and 8–10 Hz, respectively[57]. Since the zircon width in peridotite is generally 35 μm, the Hf isotope analysis of zircons from peridotite were acquired in single spot ablation mode at a spot size of 32 μm at Wuhan SampleSolution Analytical Technology Co., Ltd[58]. Experiments of *in-situ* Hf isotope ratio analysis at Wuhan SampleSolution Analytical Technology Co., Ltd. were conducted using a Neptune Plus MC-ICP-MS (Thermo Fisher Scientific, Germany) in combination with a Geolas HD excimer ArF laser ablation system (Coherent, Göttingen, Germany). Off-line selection and integration of analyte signals and mass bias calibrations were performed using ICPMSDataCal[59]. The zircon Hf isotope data are listed in Supplementary Table 6.

## In-situ Sulfur isotope analysis

A Nu plasma 1700 MC-ICP-MS (NuInstruments, UK) was combined with the RESOlution excimer laser ablation system (193 nm, RESOlution M-50, ASI) for in situ sulfur isotopic composition measurement of anhydrite and pyrite at the State Key Laboratory of Continental Evolution and Early Life, Northwestern University, China. Before the analysis, the metal minerals were well identified in the probe tablets, and the sulfides with particle size > 70 μm and smooth surface were selected for analysis. NWU-Gy and NWU-Py were the reference materials[60], and the S isotope of anhydrite and pyrite were determined with spot sizes of 51 and 32 um, respectively. The long-term (over seven months) analysis precision of the NWU-Gy, and NWU-Py LA-MC-ICP-MS measurement was 0.32‰ and 0.26‰ (2SD), respectively[60]. *In situ* sulfur isotope analyses of several pyrites were performed on a Neptune Plus MC-ICP-MS (Thermo Fisher Scientific, Bremen, Germany) equipped with a Geolas HD excimer ArF laser ablation system (Coherent, Göttingen, Germany) at the Wuhan Sample Solution Analytical Technology Co., Ltd, Hubei, China. Pyrite PPP-1 was used as a standard, and the S isotope of pyrite was determined with spot sizes of 44 μm using the single-spot ablation mode. The external precisions were 0.15–0.45‰ (2s) for $\delta^{34}S_{V\text{-}CDT}$ with an accuracy of 0.15‰ for $\delta^{34}S_{V\text{-}CDT}$[61]. Results are listed in Supplementary Table 4.

## In-situ Fe isotope analysis

Iron isotopic analysis in pyrite and magnetite was performed by 193 nm excimer ns laser ablation combined with Nu Plasma 1700 high resolution MC-ICP-MS without matrix-match calibration at the State Key Laboratory of Continental Evolution and Early Life, Northwestern University, China. Consistent lower fluence (1.5–3.5 $J \cdot cm^{-2}$) and wet plasma conditions were used to reduce the matrix effect in Fe isotopic analysis[62]. xxxx and xxxx were the reference materials, and the Fe isotope in pyrite and magnetite were determined with spot sizes of xxx and xxx um, respectively. The long-term (over seven months) analysis precision of the xxxx and xxxx LA-MC-ICP-MS measurements was XXX and XXX (2SD), respectively. Results are listed in Supplementary Table 4.

## Whole-rock C-O isotope analysis

Whole-rock carbon and oxygen isotope analyses were conducted at the State Key Laboratory of Continental Evolution and Early Life, Northwestern University, China, following the industry standard DZ/T 0184.28-2024. Sample powders (ground to 200 mesh) were loaded into reaction vials, flushed with $N_2$ to expel atmospheric air, and then reacted with anhydrous phosphoric acid at 70 °C for 2 h. The evolved $CO_2$ was analysed using a Thermo Fisher MAT 253 Plus isotope ratio mass spectrometer interfaced with a GasBench II continuous-flow gas preparation system. Analytical precision (1 standard deviation) was better than 0.06‰ for $\delta^{13}C_{V\text{-}PDB}$ and 0.08‰ for $\delta^{18}O_{V\text{-}PDB}$. Results are listed in Supplementary Table 4.

## The oxygen isotopes and triple-oxygen isotopes of anhydrite

Three hornblendite samples containing anhydrite from different depths of drill hole DH0101 were crushed. Anhydrite separates (30–40 mg) were handpicked under a binocular microscope, ground to 200 mesh in an agate mortar, and sent to the International Isotope Effect Research Center, Nanjing University, for isotopic analysis. Triple-oxygen-isotope compositions ($\Delta'^{17}O$) were measured using a MAT 253 Plus mass spectrometer; detailed analytical protocols are described in ref. 63.

Oxygen-isotope compositions ($\delta^{18}O$, reported relative to V-SMOW) of anhydrite were analysed using an EA-HT-Delta V Plus system, calibrated against the in-house standard ICIER-SO-3 ($\delta^{18}O$ = +11.81‰ versus V-SMOW). Repeated measurements of ICIER-SO-3 ($n$ = 3) yielded a reproducibility of 0.2‰ (1σ). Results are listed in Supplementary Table 4.

## The determination of Calcium isotope

One Proterozoic schist, one Proterozoic marble, two phlogopite wehrlites, two phlogopite clinopyroxenites, six hornblendites, and three anhydrite single-mineral samples were analyzed for calcium isotopes. Solid samples were precisely weighed and dissolved in 2.5 M HCl. The fractions containing Ca were collected in PFA beakers, evaporated until dryness and re-dissolved in 3.5% $HNO_3$. Calcium was isolated from the sample matrix using AG 50W-8X cation exchange resin to remove elements (i.e., Sr) that generate interferences beyond the ion optical resolving power of the mass spectrometer and to ensure that varying the matrix load from the aspirated solutions introduced into the plasma did not influence the mass bias. The solution was then introduced into multi-collector inductively coupled plasma mass spectrometry (MC-ICP-MS). All $\delta^{44/42}Ca$ values were reported in permil (‰) and calculated relative to the reference standard NIST SRM 915a[64].

Standard Deviation was calculated from two independent consecutive measurements to monitor analytical precision. The relative standard deviation (RSD) of $\delta^{42/44}Ca$ is less than 0.05%. Results are listed in Supplementary Table 4.

## The composition of parental magma

The composition of parental magma can be inferred by utilizing the olivine composition and the Mg-Fe distribution coefficient between olivine (Ol) and magmatic liquid (Liq): $K_D = X(FeO/MgO)^{Ol}/X(FeO/MgO)^{Liq} = 0.30 \pm 0.03$ (molar ratio)[65]. The most primitive olivine (i.e., Fo = 90.05) was selected from a peridotite sample at 7.7 m from DH2403. This rare sulfide feature and the low Ni content (0.16%) of this sample can prevent the Fe-Ni exchange between olivine and sulfide liquids. The FeO/MgO molar ratio of equilibrium melt coexisting with this olivine is inferred to be 0.3683 (the FeO/MgO mass ratio is 0.6565), which represents the lowest FeO/MgO ratio in the parental magma of the Langmuri deposit. To avoid the influence of surface weathering on elemental content, only drillcore samples were selected for parent magma estimation. After loss-on-ignition (LOI) correction (Supplementary Table 5), the MgO and FeO contents of the peridotite showed a negative correlation (Extended Data Fig. 14a). Therefore, a functional relationship between MgO and FeO contents in peridotites could be obtained. The values corresponding to the intersection of this functional relationship with the straight line FeO = 0.6565MgO (in wt.%) represent the MgO and FeO contents of the parent magma. The estimated composition of the parental magma is 11.49 wt.% MgO and 7.54 wt.% FeO. The MgO and $SiO_2$ contents of the rocks after LOI correction in the southern part of the Langmuri deposit show a roughly negative correlation. The $SiO_2$ content corresponding to an MgO content of 11.49 wt% roughly represents the $SiO_2$ content in the parent magma, which is 51.96 wt% (Extended Data Fig. 14b). The estimated composition for parental magma is consistent with the characteristics of high-Mg basalt[66].

## Calculation of metallic element content of 100% sulfides in ultramafic rocks

In this paper, the content of each PGE in 100% sulfide was calculated according to Equation 1 [67]:

$$C_{100\%Sul} = \frac{C_{wr} \times 100}{2.527 \times S + 0.3408 \times Cu + 0.4715 \times (Ni - Ni_{Sil})}, (1)$$

where $C_{100\%Sul}$ is the content of each metal in the 100% sulfide; $C_{wr}$ is the content (%) of the metal element in the whole rock; S, Cu, and Ni are the concentration of these elements in the whole rock; $Ni_{Sil}$ is the nickel content in silicates and oxides. For phlogopite wehrlite, $Ni_{Sil}$ is approximately equal to the nickel content in olivine. The nickel content in olivine in the Langmuri deposit is 1065 ug/g, which is the average Ni content of olivines from DH4901. For clinopyroxenite and hornblendite rocks, since there is no olivine, $Ni_{Sil}$ is assumed to be 0. The calculation process is listed in Supplementary Table 9.

## Calculation of metallic element content of 100% sulfides in porphyry copper deposits

The method for calculating the metal element content of 100% sulfides in ultramafic rocks is based on the premise that the sulfides in the rocks are pyrrhotite (Fe1-xS), pentlandite [$(Fe,Ni)_9S_8$], chalcopyrite ($FeCuS_2$)[68]. This method is obviously not applicable to porphyry Cu deposits where pyrite ($FeS_2$), chalcopyrite, bornite ($Cu_5FeS_4$), Galena (PbS), sphalerite (ZnS), and molybdenite ($MoS_2$) are the main sulfides. A new method is needed to calculate the metal element content of 100% sulfides in porphyry copper deposits. In porphyry deposit studies, only three deposits—the Skouries porphyry deposit in Greece, the Muratdere porphyry deposit in Turkey, and the El Teniente deposit in Chile—have had their PGEs, Fe, Cu, Mo, Pb, Zn, and Ni simultaneously reported[9]. These samples are all quartz vein ore. It can be roughly assumed that the Fe content of the rocks comes entirely from sulfides. Therefore, the sum of the elements S, Fe, Cu, Mo, Pb, Zn, and minor metals including Ni, Co, Ag, Au, Te, Bi, Se, Ru, Rh, Pt, Pd, Ir, and Os, represents the total mass of sulfides ($m_{Sul}$). It is important to note that samples containing sulfate minerals were removed to ensure that the whole-rock S content represents the S content in sulfides as accurately as possible. Furthermore, samples with S content less than 0.5 wt% were excluded from the calculation to minimize errors. Multiplying individual metal element content by 100 and then dividing by $m_{Sul}$ gives the content of this metal element in 100% sulfide.

$$C_{100\%Sul} = \frac{C_{wr} \times 100}{C_S + C_{Fe} + C_{Cu} + C_{Mo} + C_{Pb} + C_{Zn} + C_{minor\ metal}}$$

where $C_{100\%Sul}$ is the content of each metal in the 100% sulfide; $C_{wr}$ is the content (%) of the metal element in the whole rock; $C_S$, $C_{Fe}$, $C_{Cu}$, $C_{Mo}$, $C_{Pb}$, $C_{Zn}$, and $C_{minor\ metal}$ are the concentration of these elements in the whole rock. The calculation process is listed in Supplementary Table 10.

## Sulfide segregation degree calculation

The sulfide liquid/silicate melt partition coefficients ($D^{Sul/Sil}$) of Cu ($10^2$ to n × $10^3$) are much lower than those of Pd ($10^4$ to $10^5$), which indicates that the Cu/Pd ratio will strongly increase once sulfide liquids segregate from the silicate magma[42,43,69]. The lowest Cu/Pd ratio of Langmuri ore rocks is 2021. This values is much lower than the primitive mantle Cu/Pd ratio (7692)[70], suggesting that the parent magma of Langmuri never experienced sulfide saturation.

Metal contents of sulfide liquids are a function of their abundances in parental silicate magmas, sulfide melt/silicate liquid partition coefficients, and the R-factor[71]:

$$C_i^{Sul} = \frac{C_i^{Sil} \times D_i \times (R+1)}{R + D_i},$$

where $C_i^{sul}$ and $C_i^{sil}$ represent the concentrations of element i in the sulfide melt and in the parental silicate magma, respectively; $D_i$ is the sulfide melt/silicate liquid partition coefficient of element i; and R is the ratio of the mass of silicate magma to the mass of sulfide that reached equilibrium together.

The Langmuri parental magma is assumed to be similar to high-Mg basaltic composition, which has 14 ppb Pd[72] and 300 ppm Ni[73,74]. The partition coefficients of Pd and Ni between sulfide melt and silicate magma used in this calculation are 100,000 and 800, respectively[42]. The calculation process is listed in Supplementary Table 11. The right dashed line in Fig. 4 displays sulfide liquids separated from PGE-weakly depleted magma with 1.89 ppb Pd100% and 295.2 ppm Ni100% (because of a 0.002% sulfide removal from PGE-rich magma) under different R values. The right dashed-dotred line in Figure 4 shows the sulfide liquids produced by sulfide immiscibility in a heavily PGE-depleted magma with 0.094 ppb Pd100% and 288.3 ppm Ni100% (because of a 0.005% sulfide removal from PGE-rich magma). Some Pd and Ni contents of the Langmuri sulfides plot above or along the no sulfide removal dashed line model 1 in Fig. 4b, indicating that the parent magma of Langmuri was no sulfide saturated. The metal contents in residual magmas after an earlier sulfide removal can be calculated on the basis of the Rayleigh Law:

$$\frac{C_i^L}{C_i^o} = (1 - F_{Sul})^{(D_i - 1)},$$

in which $C_i^O$ and $C_i^L$ represent abundances of element i in the original and residual magmas, respectively, $D_i$ is the partition coefficient, and $F_{Sul}$ is the mass fraction of segregated sulfide.

## The calculation of Ni and Pd contents in the residual magma during the olivine fractionation crystallization process of high-Mg basaltic magma (parent magma A)

The elemental content of residual magma during olivine fractionation crystallization of high-Mg basaltic magma (parent magma A) can be calculated using the Rayleigh Law:

$$\frac{C_i^L}{C_i^o} = (1 - F_{Ol})^{(D_i - 1)},$$

in which $C_i^O$ and $C_i^L$ represent abundances of element i in the original and residual magmas, respectively, $D_i$ is the partition coefficient, and $F_{Ol}$ is the mass fraction of olivine. The partition coefficients of Pd and Ni between olivine and silicate melt used in this calculation are 0.03 and 11, respectively[75-77]. The calculation process is listed in Supplementary Table 12.

# Acknowledgements

We thank the financial support from the Deep Earth Probe and Mineral Resources Exploration—National Science and Technology Major Project (grant numbers 2025ZD1007100 and 2025ZD1007106), the National Key Research and Development Program Project (2025YFF0517004), the Science and Technology Innovation Project of Laoshan National Laboratory (LSKJ202502704), and the New Star of South China Sea Talent Project (NHXXRCXM202339).

# Author contributions

1 Holland, H. D. 100th Anniversary Special Paper: Sedimentary Mineral Deposits and the Evolution of Earth's Near-Surface Environments. *Economic Geology* **100**, 1489–1509, doi:10.2113/gsecongeo.100.8.1489 (2005).

2 Moreira, H. *et al.* Sub-arc mantle fugacity shifted by sediment recycling across the Great Oxidation Event. *Nat. Geosci.* **16**, 922–927 (2023).

3 Tostevin, R. & Mills, B. J. W. Reconciling proxy records and models of Earth's oxygenation during the Neoproterozoic and Palaeozoic. *Interface Focus* **10**, 20190137, doi:10.1098/rsfs.2019.0137 (2020).

4 Wang, H. *et al.* Two-billion-year transitional oxygenation of the Earth's surface. *Nature* **645**, 665–671, doi:10.1038/s41586-025-09471-4 (2025).

5 Zhu, X.-X., Duan, W.-Y., Gerya, T., Zhou, X. & Tian, J.-C. The Mantle Fe3+/ΣFe Ratio Has Doubled Since the Early Archean. *Nature Communications* **17**, 429, doi:10.1038/s41467-025-66969-1 (2026).

6 Stolper, D. A. & Keller, C. B. A record of deep-ocean dissolved O2 from the oxidation state of iron in submarine basalts. *Nature* **553**, 323–327, doi:10.1038/nature25009 (2018).

7 Song, X. Y. *et al.* The Giant Xiarihamu Ni-Co Sulfide Deposit in the East Kunlun Orogenic Belt, Northern Tibet Plateau, China. *Economic Geology* **111**, 29–55 (2016).

8 Singer, D. A., Berger, V. I. & Moring, B. C. Porphyry copper deposits of the world: Database and grade and tonnage models. *US Geological Survey Open-File Report 2008-1155* (2008).

9 McFall, K., McDonald, I. & Wilkinson, J. J. in *Tectonomagmatic Influences on Metallogeny and Hydrothermal Ore Deposits: A Tribute to Jeremy P. Richards (Volume II)* Vol. 2 (eds Ali

Sholeh & Rui Wang) 277–295 (Society of Economic Geologists, 2021).

10 Holwell, D. A. *et al.* A metasomatized lithospheric mantle control on the metallogenic signature of post-subduction magmatism. *Nature Communications* **10**, 3511 (2019).

11 Bénard, A. *et al.* Oxidising agents in sub-arc mantle melts link slab devolatilisation and arc magmas. *Nature Communications* **9**, 3500, doi:10.1038/s41467-018-05804-2 (2018).

12 Wirth, R., Kaminsky, F., Matsyuk, S. & Schreiber, A. Unusual micro-and nano-inclusions in diamonds from the Juina Area, Brazil. *Earth and Planetary Science Letters* **286**, 292–303 (2009).

13 Zelenski, M. *et al.* Sulfide-sulfate metasomatism and nickel release in the suprasubduction mantle. *Earth and Planetary Science Letters* **626**, 118500 (2024).

14 Zhang, Z. M. *et al.* Fluids in deeply subducted continental crust: petrology, mineral chemistry and fluid inclusion of UHP metamorphic veins from the Sulu orogen, eastern China. *Geochimica Et Cosmochimica Acta* **72**, 3200–3228 (2008).

15 Mapiloko, M. *et al.* Contribution of the oldest Paleoproterozoic marine sulfate evaporites to Bushveld Complex Lower Zone mineralization. *Geology* **53**, 355–359, doi:10.1130/g52550.1 (2025).

16 Ripley, E. M., Lightfoot, P. C., Li, C. & Elswick, E. R. Sulfur isotopic studies of continental flood basalts in the Noril'sk region: implications for the association between lavas and ore-bearing intrusions. *Geochimica Et Cosmochimica Acta* **67**, 2805–2817 (2003).

17 Audétat, A., Chang, J. & Gaynor, S. P. Widespread anhydrite saturation in Laramide-age arc magmas of the southwestern USA. *Geology* **54**, 19–23 (2026).

18 Fu, L. *et al.* Growth of early Paleozoic continental crust linked to the Proto-Tethys subduction and continental collision in the East Kunlun Orogen, northern Tibetan Plateau. *Geol. Soc. Am. Bull.* **135**, 1709–1733 (2023).

19 Liu, B. *et al.* Early Paleozoic tectonic transition from ocean subduction to collisional orogeny in the Eastern Kunlun region: Evidence from Huxiaoqin Mafic rocks. *Acta Petrologica Sinica* **29**, 2093–2106 (2013).

20 Liu, Y. G. *et al.* The Dynamic Sulfide Saturation Process and a Possible Slab Break-off Model for the Giant Xiarihamu Magmatic Nickel Ore Deposit in the East Kunlun Orogenic Belt, Northern Qinghai-Tibet Plateau, China. *Economic Geology* **113**, 1383–1417, doi:https://doi.org/10.5382/econgeo.2018.4596 (2018).

21 Ridolfi, F. Amp-TB2: An Updated Model for Calcic Amphibole Thermobarometry. *Minerals* **11**, 324 (2021).

22 Schauble, E. A., Rossman, G. R. & Taylor, H. P. Theoretical estimates of equilibrium Fe-isotope fractionations from vibrational spectroscopy. *Geochimica Et Cosmochimica Acta* **65**, 2487–2497, doi:https://doi.org/10.1016/S0016-7037(01)00600-7 (2001).

23 Claypool, G. E., Holser, W. T., Kaplan, I. R., Sakai, H. & Zak, I. The age curves of sulfur and oxygen isotopes in marine sulfate and their mutual interpretation. *Chemical Geology* **28**, 199–260, doi:https://doi.org/10.1016/0009-2541(80)90047-9 (1980).

24 Yang, X.-M. & Le Bas, M. J. Chemical compositions of carbonate minerals from Bayan Obo, Inner Mongolia, China: implications for petrogenesis. *Lithos* **72**, 97–116, doi:https://doi.org/10.1016/j.lithos.2003.09.002 (2004).

25 Kang, J. *et al.* Calcium isotopic fractionation in mantle peridotites by melting and metasomatism and Ca isotope composition of the Bulk Silicate Earth. *Earth and Planetary*

*Science Letters* **474**, 128–137 (2017).

26 Amsellem, E. *et al.* Calcium isotopic evidence for the mantle sources of carbonatites. *Science advances* **6**, eaba3269, doi:doi:10.1126/sciadv.aba3269 (2020).

27 Chen, C. *et al.* Calcium isotope fractionation during magmatic processes in the upper mantle. *Geochimica Et Cosmochimica Acta* **249**, 121–137 (2019).

28 Xiao, Z.-C., Zhou, C., Kang, J.-T., Wu, Z.-Q. & Huang, F. The factors controlling equilibrium inter-mineral Ca isotope fractionation: Insights from first-principles calculations. *Geochimica Et Cosmochimica Acta* **333**, 373–389 (2022).

29 Hu, Q. *et al.* $FeO_2$ and FeOOH under deep lower-mantle conditions and Earth's oxygen–hydrogen cycles. *Nature* **534**, 241–244 (2016).

30 Holland, H. D. Volcanic gases, black smokers, and the great oxidation event. *Geochimica Et Cosmochimica Acta* **66**, 3811–3826, doi:https://doi.org/10.1016/S0016-7037(02)00950-X (2002).

31 Cottrell, E., Canil, D., Langmuir, C., Evans, K. A. & Gaillard, F. Earth's past and present mantle oxygen fugacity. *Nature Reviews Earth & Environment* **6**, 728–746, doi:10.1038/s43017-025-00735-1 (2025).

32 Liu, Y. *et al.* Oldhamite: A new link in upper mantle for C-O-S-Ca cycles and an indicator for planetary habitability. *National Science Review* **10**, nwad159 (2023).

33 Schröder, S., Beukes, N. J. & Armstrong, R. A. Detrital zircon constraints on the tectonostratigraphy of the Paleoproterozoic Pretoria Group, South Africa. *Precambrian Research* **278**, 362–393, doi:https://doi.org/10.1016/j.precamres.2016.03.016 (2016).

34 Chen, Y. Evidences for the Catastrophe in Geologic Environment at about 2300 Ma and the Discussions on Several Problems. *Journal of Stratigraphy* **14**, 178–186, doi:https://doi.org/10.1016/j.gsf.2013.02.003 (1990).

35 Shields-Zhou, G. & Och, L. The case for a Neoproterozoic oxygenation event: geochemical evidence and biological consequences. *GSA Today* **21**, 4–11 (2011).

36 Lyons, T. W. & Gill, B. C. Ancient Sulfur Cycling and Oxygenation of the Early Biosphere. *Elements* **6**, 93–99, doi:10.2113/gselements.6.2.93 (2010).

37 Planavsky, N. J., Bekker, A., Hofmann, A., Owens, J. D. & Lyons, T. W. Sulfur record of rising and falling marine oxygen and sulfate levels during the Lomagundi event. *Proceedings of the National Academy of Sciences* **109**, 18300–18305, doi:doi:10.1073/pnas.1120387109 (2012).

38 Ezad, I. S. *et al.* Incipient carbonate melting drives metal and sulfur mobilization in the mantle. *Science advances* **10**, eadk5979 (2024).

39 Kempe, Y. *et al.* Redox state of the deep upper mantle recorded by nickel-rich diamond inclusions. *Nat. Geosci.* **18**, 1048–1055, doi:10.1038/s41561-025-01791-4 (2025).

40 Thiéblemont, D., Stein, G. & Lescuyer, J.-L. Gisements épithermaux et porphyriques: la connexion adakite. *Comptes Rendus de l'Académie des Sciences - Series IIA - Earth and Planetary Science* **325**, 103–109, doi:https://doi.org/10.1016/S1251-8050(97)83970-5 (1997).

41 Hou, Z. Q., Mo, X., Gao, Y., Qu, X. & Meng, X. Adakite, A Possible Host Rock for Porphyry Copper Deposits: Case Studies of Porphyry Copper Belts in Tibetan Plateau and in Northern Chile. *Minerals Deposits*, 1–12, doi:10.16111/j.0258-7106.2003.01.001 (2003).

42 Mungall, J. E. & Brenan, J. M. Partitioning of platinum-group elements and Au between sulfide liquid and basalt and the origins of mantle-crust fractionation of the chalcophile elements. *Geochimica Et Cosmochimica Acta* **125**, 265–289 (2014).

43 Li, Y. & Audétat, A. Partitioning of V, Mn, Co, Ni, Cu, Zn, As, Mo, Ag, Sn, Sb, W, Au, Pb, and Bi between sulfide phases and hydrous basanite melt at upper mantle conditions. *Earth and Planetary Science Letters* **355**, 327–340 (2012).

44 Mao, Y. J., Qin, K. Z., Li, C. & Tang, D. M. A modified genetic model for the Huangshandong magmatic sulfide deposit in the Central Asian Orogenic Belt, Xinjiang, western China. *Mineralium Deposita* **50**, 65–82 (2015).

45 Canali, A., Brenan, J. & Sullivan, N. Solubility of platinum-arsenide melt and sperrylite in synthetic basalt at 0.1 MPa and 1200° C with implications for arsenic speciation and platinum sequestration in mafic igneous systems. *Geochimica Et Cosmochimica Acta* **216**, 153–168 (2017).

46 Davidson, J., Turner, S., Handley, H., Macpherson, C. & Dosseto, A. Amphibole “sponge” in arc crust? *Geology* **35**, 787–790 (2007).

47 Atherton, M. P. & Petford, N. Generation of sodium-rich magmas from newly underplated basaltic crust. *Nature* **362**, 144–146, doi:10.1038/362144a0 (1993).

48 Chung, S.-L. *et al.* Adakites from continental collision zones: Melting of thickened lower crust beneath southern Tibet. *Geology* **31**, 1021–1024, doi:10.1130/g19796.1 (2003).

49 Defant, M. J. & Drummond, M. S. Derivation of some modern arc magmas by melting of young subducted lithosphere. *Nature* **347**, 662–665, doi:10.1038/347662a0 (1990).

50 Green, T. H., Blundy, J. D., Adam, J. & Yaxley, G. M. SIMS determination of trace element partition coefficients between garnet, clinopyroxene and hydrous basaltic liquids at 2–7.5 GPa and 1080–1200°C. *Lithos* **53**, 165–187, doi:https://doi.org/10.1016/S0024-4937(00)00023-2 (2000).

51 Pertermann, M., Hirschmann, M. M., Hametner, K., Günther, D. & Schmidt, M. W. Experimental determination of trace element partitioning between garnet and silica-rich liquid during anhydrous partial melting of MORB-like eclogite. *Geochem. Geophys. Geosyst.* **5**, doi:https://doi.org/10.1029/2003GC000638 (2004).

52 Qi, L. *et al.* An improved digestion technique for determination of platinum group elements in geological samples. *Journal of Analytical Atomic Spectrometry* **26**, 1900–1904 (2011).

53 Liu, Y. S. *et al.* In situ analysis of major and trace elements of anhydrous minerals by LA-ICP-MS without applying an internal standard. *Chemical Geology* **257**, 34–43 (2008).

54 Jochum, K. P. *et al.* Determination of reference values for NIST SRM 610–617 glasses following ISO guidelines. *Geostand. Geoanal. Res.* **35**, 397–429 (2011).

55 Wu, S. *et al.* The preparation and preliminary characterisation of three synthetic andesite reference glass materials (ARM-1, ARM-2, ARM-3) for in situ microanalysis. *Geostand. Geoanal. Res.* **43**, 567–584 (2019).

56 Paton, C., Hellstrom, J., Paul, B., Woodhead, J. & Hergt, J. Iolite: Freeware for the visualisation and processing of mass spectrometric data. *Journal of Analytical Atomic Spectrometry* **26**, 2508–2518, doi:10.1039/c1ja10172b (2011).

57 Yuan, H.-L. *et al.* Simultaneous determinations of U–Pb age, Hf isotopes and trace element compositions of zircon by excimer laser-ablation quadrupole and multiple-collector ICP-MS. *Chemical Geology* **247**, 100–118 (2008).

58 Hu, Z. *et al.* “Wave” signal-smoothing and mercury-removing device for laser ablation quadrupole and multiple collector ICPMS analysis: application to lead isotope analysis. *Anal. Chem.* **87**, 1152–1157 (2015).

59 Liu, Y. S. *et al.* Continental and Oceanic Crust Recycling-induced Melt-Peridotite Interactions in the Trans-North China Orogen: U-Pb Dating, Hf Isotopes and Trace Elements in Zircons from Mantle Xenoliths. *Journal of Petrology* **51**, 537–571 (2010).

60 Peng, D. *et al.* Three new potential sulfur reference materials (pyrite, gypsum, and arsenopyrite) for in situ sulfur isotope analysis by laser ablation MC-ICP-MS. *Journal of Analytical Atomic Spectrometry* **39**, 2235–2244 (2024).

61 Fu, J. *et al.* In situ sulfur isotopes ($\delta^{34}S$ and $\delta^{33}S$) analyses in sulfides and elemental sulfur using high sensitivity cones combined with the addition of nitrogen by laser ablation MC-ICP-MS. *Analytica Chimica Acta* **911**, 14–26 (2016).

62 Chen, K., Yuan, H., Bao, Z. a. & Lv, N. Accurate analysis of Fe isotopes in Fe- dominated minerals by excimer laser ablation MC-ICP-MS on wet plasma conditions. *At. Spectrosc.* **42**, 282–293 (2021).

63 Wei, Y., Yan, H., Peng, Y. & Bao, H. Quantitative conversion of sulfate oxygen for high-precision triple oxygen isotope analysis. *Anal. Chem.* **96**, 19387-19395 (2024).

64 Eisenhauer, A. *et al.* Proposal for International Agreement on Ca Notation Resulting from Discussions at Workshops on Stable Isotope Measurements Held in Davos (Goldschmidt 2002) and Nice (EGS-AGU-EUG 2003). *Geostand. Geoanal. Res.* **28**, 149–151, doi:https://doi.org/10.1111/j.1751-908X.2004.tb01051.x (2004).

65 Roeder, P. & Reynolds, I. Crystallization of chromite and chromium solubility in basaltic melts. *Journal of Petrology* **32**, 909–934 (1991).

66 Sun, S. S., Nesbitt, R. W. & Mcculloch, M. T. Geochemistry and petrogenesis of Archaean and early Proterozoic siliceous high-magnesian basalts. *in Crawfors, A.J., ed., Boninites and related rocks: Winchester, Massachusetts, Unwin Hyman, Boninites and related rocks*, 149–173 (1989).

67 Barnes, S. J. & Ripley, E. M. Highly siderophile and strongly chalcophile elements in magmatic ore deposits. *Rev. Mineral. Geochem.* **81**, 725–774 (2016).

68 Li, C., Maier, W. & De Waal, S. Magmatic Ni-Cu versus PGE deposits: Contrasting genetic controls and exploration implications. *S. Afr. J. Geol.* **104**, 309–318 (2001).

69 Bennett, V., Norman, M. & Garcia, M. Rhenium and platinum group element abundances correlated with mantle source components in Hawaiian picrites: sulphides in the plume. *Earth and Planetary Science Letters* **183**, 513–526 (2000).

70 McDonough, W. F. & Sun, S. S. The composition of the Earth. *Chemical geology* **120**, 223–253 (1995).

71 Campbell, I. & Naldrett, A. The influence of silicate: sulfide ratios on the geochemistry of magmatic sulfides. *Economic Geology* **74**, 1503–1506 (1979).

72 Hamlyn, P. R., Keays, R. R., Cameron, W. E., Crawford, A. J. & Waldron, H. M. Precious metals in magnesian low-Ti lavas: Implications for metallogenesis and sulfur saturation in primary magmas. *Geochimica Et Cosmochimica Acta* **49**, 1797–1811 (1985).

73 Li, C., Xu, Z., de Waal, S. A., Ripley, E. M. & Maier, W. D. Compositional variations of olivine from the Jinchuan Ni-Cu sulfide deposit, western China: implications for ore genesis. *Mineralium Deposita* **39**, 159–172 (2004).

74 Zhao, J. H. & Zhou, M. F. Neoproterozoic high-Mg basalts formed by melting of ambient mantle in South China. *Precambrian Research* **233**, 193–205 (2013).

75 Puchtel, I. S. & Humayun, M. Platinum group element fractionation in a komatiitic basalt lava lake. *Geochimica Et Cosmochimica Acta* **65**, 2979–2993 (2001).

76 Pedersen, A. K. Basaltic glass with high-temperature equilibrated immiscible sulphide bodies with native iron from Disko, central West Greenland. *Contributions to Mineralogy and Petrology* **69**, 397–407 (1979).

77 Nabelek, P. I. Nickel partitioning between olivine and liquid in natural basalts: Henry's Law behavior. *Earth and Planetary Science Letters* **48**, 293–302 (1980).